\documentclass[aps,prd,nofootinbib,onecolumn,preprintnumbers,floatfix,superscriptaddress,bibliography]{revtex4-2}
\usepackage[colorlinks, linkcolor=red, anchorcolor=blue, citecolor=blue, colorlinks=true, urlcolor=blue]{hyperref}
\usepackage{amsmath}
\usepackage{graphicx}
\usepackage{fontawesome}
\usepackage{ulem}
\usepackage{bm}
\usepackage{xspace}
\usepackage{amssymb}
\usepackage{gensymb}
\usepackage{multirow}
\usepackage{threeparttable}
\usepackage{amsthm}
\usepackage{mathrsfs}
\usepackage{indentfirst}
\usepackage{subfigure}
\usepackage[figuresright]{rotating}
\usepackage{epstopdf}
\usepackage{breakurl}
\usepackage{lipsum}
\usepackage{cancel}
\usepackage{slashed}
\usepackage{orcidlink}

\begin{document}

\title{Probing Memory-Burdened Primordial Black Holes with Galactic Sources observed by LHAASO}

\author{Xiu-Hui Tan}
\affiliation{Key Laboratory of Theoretical Physics, Institute of Theoretical Physics, Chinese Academy of Sciences, Beijing 100190, China}
\email{tanxh@itp.ac.cn}

\author{Yu-Feng Zhou}
\affiliation{Key Laboratory of Theoretical Physics, Institute of Theoretical Physics, Chinese Academy of Sciences, Beijing 100190, China}
\affiliation{School of Physical Sciences, University of Chinese Academy of Sciences, Beijing 100049, China}
\affiliation{School of Fundamental Physics and Mathematical Sciences, Hangzhou Institute for Advanced Study, UCAS, Hangzhou 310024, China}
\affiliation{International Centre for Theoretical Physics Asia-Pacific, Beijing/Hangzhou, China}
\email{yfzhou@itp.ac.cn}

\date{\today}

\begin{abstract}
	The recently identified \textit{memory burden} effect has the potential to significantly decelerate the evaporation of black holes. Specifically, when approximately half of a black hole's initial mass has been radiated away, the evaporation process is halted. This mechanism allows very light primordial black holes (PBHs) with masses $m_{\rm PBH}<10^{15}$ g to persist until the present day and may contribute to the dark matter (DM) content of the universe. In this work, we focus on PBHs with masses $\lesssim 10^{9}$ g. Due to the memory burden effect, these PBHs emit high-energy gamma-rays, which in turn alter the corresponding observed energy spectra. To investigate the constraints on the masses and DM abundance of PBHs, we analyze data from four Galactic sources measured by the Large High Altitude Air Shower Observatory (LHAASO), including the Crab Nebula, LHAASO J2226+6057, LHAASO J1908+0621, and LHAASO J1825-1326. Our findings indicate that the ultra-high-energy gamma-ray spectra from these Galactic sources provide crucial probes for light PBHs, thereby significantly constraining their potential contribution to DM.
\end{abstract}

\maketitle

\section{Introduction}\label{sec:intro}
Dark matter (DM) is one of the most crucial components of our Universe and has been the subject of extensive research for decades, with numerous candidate particles proposed. Recent studies have suggested that primordial black holes (PBHs), which are the hypothetical black holes that may have formed in the early Universe, could be a viable candidate for DM.
PBHs are expected to emit standard model particles through the process of Hawking evaporation (HE) \cite{1975CMaPh..43..199H, 10.1093/mnras/152.1.75}. The potential of asteroid-mass PBHs as DM has been explored in many studies through gamma-ray, neutrino emission, and cosmic-ray observations \cite{Bambi:2008kx, DeRocco:2019fjq, Laha:2019ssq, laha_integral_2020, Iguaz:2021irx, Chen:2021ngo, Tan:2022lbm, Korwar:2023kpy, Tan:2024nbx, Huang:2024xap}. However, it has long been believed that PBHs with masses smaller than $\sim 10^{15}{~\rm g}$ would have completely evaporated within the current age of the Universe. This conclusion is based on Hawking's original analysis, which neglected the influence of the emitted radiation on the quantum state of the black hole.

Recent work \cite{Dvali:2018xpy, Dvali:2020wft, Dvali:2024hsb, Haque:2024eyh} has claimed that the information content within a PBH could slow down its evaporation through the effect of \textit{memory burden}. Due to this effect, the evolution of PBHs is divided into two stages. Initially, they evaporate according to the ordinary Hawking formula. At approximately half of their lifetime $t_{\rm half}$, they transition to a second stage. Although the exact behavior after this time remains unclear, the system becomes effectively stabilized and gains a much longer lifetime \cite{Dvali:2020wft, Dvali:2021tez}.
One of the two possible scenario have been proposed in reference \cite{Dvali:2020wft}, it implied the lifetime of a PBH could be prolonged substantially. This has studied both in numerically and analytically and has significant phenomenological consequences in the present Universe, such as big bang nucleosynthesis (BBN) and the cosmic microwave background (CMB). Particularly it reopens the lower masses of PBHs lighter than $10^{15}{~\rm g}$, and extending the viable mass range down to $\sim 10^4{~\rm g}$.

Light PBHs contribute to the DM abundance and possess extremely small event horizons. They are typically assumed to be initially Poisson-distributed in the early Universe following their formation. Subsequently, like other DM candidates, they cluster into DM halos. This clustering process results in local PBH and DM densities that are significantly higher than the cosmic average, and their distribution could follow the DM halo profile.
The presence of PBHs near the galactic center has the potential to induce occasional contributions to the gamma-ray spectrum. Previous constraints derived from high-energy diffuse gamma-ray flux measurements by many detectors, below $10^6$ GeV, have imposed significant bounds on the properties of light PBHs \cite{Alexandre:2024nuo, Thoss:2024hsr, Chianese:2024rsn, Dvali:2025ktz, Montefalcone:2025akm, Liu:2025vpz, Zantedeschi:2024ram}.
At energies higher than $\sim 10^5$ GeV, gamma-rays can interact with galactic and extragalactic background radiation, leading to the kinematically possible creation of electron-positron pairs. This interaction results in the attenuation of the gamma-ray flux as it propagates to Earth. Consequently, light PBHs within the local galaxy modify the expected gamma-ray spectrum at high energies within the standard astrophysical framework.

Notably, the Large High Altitude Air Shower Observatory (LHAASO) \cite{Ma:2022aau} is one of the most significant observatories globally, having made remarkable contributions to the measurement of high energy gamma-ray spectra in this field. Utilizing two detectors, the Water Cherenkov Detector Array (WCDA) and the Kilometer Square Array (KM2A), the LHAASO collaboration has provided several gamma-ray spectra that span more than three orders of magnitude in energy. These spectra include those from the Crab Nebula and 12 other ultra-high-energy gamma-ray extended Galactic sources \cite{LHAASO:2021gok}, with precise sensitivity. However, except for the Crab Nebula, these sources have not yet been firmly localized and identified, and the extreme acceleration processes remain unknown.

In this work, we analyze prominent measurements of high-energy gamma-ray spectra from LHAASO. We focus on four sources: the Crab Nebula, LHAASO J226+6057, LHAASO J1908+0621, and LHAASO J1825-1326, to impose constraints on the parameter space of light PBHs, as also discussed in \cite{Li:2024ivs}. The gamma-ray emissions from these Galactic sources indicate the presence of active or recent particle accelerators within or in the vicinity of the gamma-ray emitting regions. This suggests that high-energy gamma-ray spectra can serve as promising probes to investigate the distribution of PBHs in the Galactic center.

The paper is organized as follows: we briefly discuss the memory burden effect and the modified gamma-ray spectra resulting from this process in Section \ref{sec:MB}. Then, in Section \ref{sec:ana}, we present and discuss our results. Our main conclusions are summarized in Section \ref{sec:concl}.

\section{Memory Burden effect}\label{sec:MB}
In this section, we briefly introduce the high-energy gamma-ray emission from PBHs due to evaporation and the modifications resulting from the memory burden effect.
According to Hawking's semiclassical calculation, a PBH with mass $M_{\rm PBH}$ should emit particles with a nearly thermal spectrum, peaked at the Hawking temperature, which is defined as:
\begin{eqnarray}
	T_{\rm H} = \frac{\hbar c^3}{8 \pi G M_{\rm PBH}}\simeq 10^4 \left(\frac{10^9 {~\rm g}}{M_{\rm PBH}}\right){~\rm GeV},
\end{eqnarray}
where $\hbar$ is the reduced Planck constant, $c$ is the speed of light, and $G$ is the gravitational constant. This temperature characterizes the thermal nature of the emitted radiation, which includes gamma rays among other particles.
In conventional scenario, the evaporation process consume the entire PBH mass. During this process, PBH entropy can be written as
\begin{eqnarray}
	S(M_{\rm PBH}) = \frac{4 \pi M_{\rm PBH}^2 G k_B}{\hbar c}\approx 2.6\times 10^{10} k_B \left( \frac{M_{\rm PBH}}{1{~\rm g}} \right)^2,
\end{eqnarray}
where $k_B$ is Boltzmann constant. 

But when memory burden happens, this duration at the end of the semiclassical phase resists the evaporation with a fraction $q=1/2$ as the memory burden when the black hole has lost half of its mass.
\begin{eqnarray}
	t_q = \tau_{\rm PBH}(1-q^3)\simeq 4.4 \times 10^{17} \left(\frac{M_{\rm PBH}}{10^{15}{~\rm g}}\right)^3{~\rm s},
\end{eqnarray}
with $\tau_{\rm PBH}$ the time of complete evaporation and the memory burden mass is correspond to 
\begin{eqnarray}
	M_{\rm PBH}^{\rm mb}=qM_{\rm PBH}.
\end{eqnarray}
Thus, when $t<t_q$, the stored information on the PBH event horizon induces a back-reaction and slows down the decay rate by a certain negative power of the PBH entropy
\begin{eqnarray}
	\frac{{\rm d}M_{\rm PBH}^{\rm mb}}{{\rm d}t}=\frac{1}{S(M_{\rm PBH})^k}\frac{{\rm d}M_{\rm PBH}}{{\rm d}t}, k>0.
\end{eqnarray}
Then the gamma-rays from a non-rotating, neutral PBH of mass $M_{\rm PBH}$ at memory burden process is 
\begin{eqnarray}
	\frac{{\rm d}^2N_\gamma^{\rm mb}}{{\rm d}E{\rm d}t}=S(M_{\rm PBH})^{-k}\frac{{\rm d}^2N_\gamma}{{\rm d}E{\rm d}t},
\end{eqnarray}
where ${{\rm d}^2N_\gamma}/{{\rm d}E{\rm d}t}$ is the semi-classical gamma-ray emission rate, which is given by 
\begin{eqnarray}
	\frac{{\rm d}^2N_\gamma}{{\rm d}E{\rm d}t} = \frac{g_{\gamma}}{2\pi}\frac{\mathcal{F}(E, M_{\rm PBH})}{e^{E/T_{\rm H}}-1},
\end{eqnarray}
where $g_\gamma=2$ denotes the internal degrees of freedom of the photons, and $\mathcal{F}(E, M_{\rm PBH})$ is the gray-body factor. The primary spectrum and secondary emission from hadronization of light PBHs are expected, here we implement the spectra computed by \texttt{BlackHawk 2.3} \cite{Arbey:2019mbc, Arbey:2021mbl} and also \texttt{HDMSpectra} \cite{Bauer:2020jay} for high-energy gamma-ray emission. Then the flux of a certain of Galactic Sources (GS) from memory burden PBHs takes the following expression:
\begin{eqnarray}
	\frac{{\rm d}^2\phi_{\gamma}^{\rm GS}}{{\rm d}E_\gamma {\rm d}t}=\frac{f_{\rm PBH}}{4\pi M^{\rm mb}_{\rm PBH}} \frac{{\rm d}^2N^{\rm mb}_\gamma}{{\rm d}E{\rm d}t} \mathcal{D}(E_\gamma, \Delta \Omega),
\end{eqnarray}
the differential gamma-ray flux from PBHs depends on the arrival directions, morphology, and deposited energies of the events. Concurrently, interactions between photons and background radiation can attenuate the measured gamma-rays. Therefore, the relevant quantity is encapsulated in the $\mathcal{D}$ factor, which is integrated over the region of interest (ROI) along the line of sight (l.o.s). This factor is given by:
\begin{eqnarray}
    \mathcal{D}(E_\gamma, \Delta \Omega)=\frac{1}{\Delta \Omega} \int_{\rm ROI}{\rm d}\Omega \int_{\rm l.o.s} {\rm d}s \rho_{\rm DM}(r) e^{-\tau_{\gamma\gamma}(E_\gamma, s, b, \ell)},
    \label{equ:Dfact}
\end{eqnarray}
where $\Omega$ is the solid angle of the ROI, and $\rho_{\rm DM}$ represents the dark matter halo profile with galactocentric radius $r = \sqrt{s^2 + R_\odot^2-2s R_\odot \cos{b}\cos{\ell}}$. Since the signals from PBHs are involved across the Milky Way, the peaks toward the galactic center and along the its plane, so $\Delta \Omega=4 \pi$ to average the $\mathcal{D}$ factor.

We adopt the classical Navarro-Frenk-White (NFW) profile \cite{Navarro_1997}, given by $\rho_{\rm NFW}(r) = \rho_s \frac{r_s}{r}\left( 1+\frac{r}{r_s} \right)^{-2}$, where $\rho_s=0.23{\rm ~GeV~cm^{-3}}$, $r_s=9.98{~\rm kpc}$. The distance from the solar position to the Galactic center is $R_\odot = 8.5{~\rm kpc}$.
In eqnarray \ref{equ:Dfact}, the optical depth $\tau_{\gamma \gamma}$ for Galactic sources from LHAASO is adopted from the values provided in reference \cite{LHAASO:2021gok}, corresponding to specific scenarios for different samples.
Within the inner region of LHAASO, the diffuse emission is confined to the range $15^\circ <\ell < 125^\circ$ and $|b|<5^\circ$. Light-mass PBHs produce higher energy gamma-ray emissions due to the distribution of PBHs in the Galaxy. Therefore, the Galactic sources from LHAASO correspond to PBH masses in the range $10^4-10^{9}{~\rm g}$. We integrate the l.o.s. distances for each Galactic source, which are taken from Extended Data Table. 2 in reference \cite{LHAASO:2021gok}.

\begin{figure}
	\centering
	\includegraphics[width=0.49\textwidth]{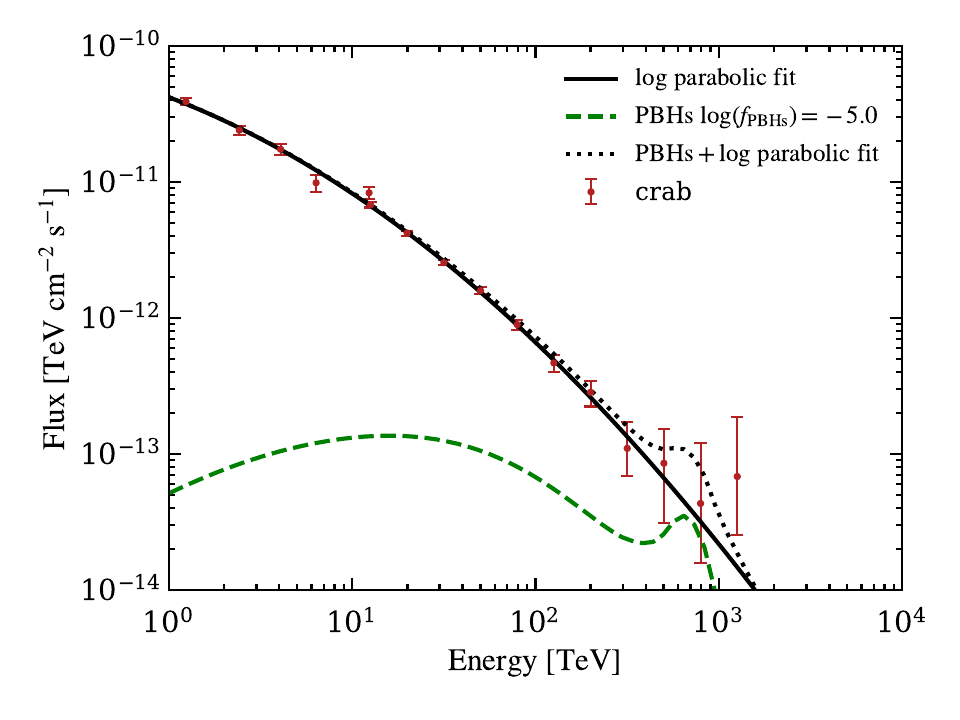}
	\includegraphics[width=0.49\textwidth]{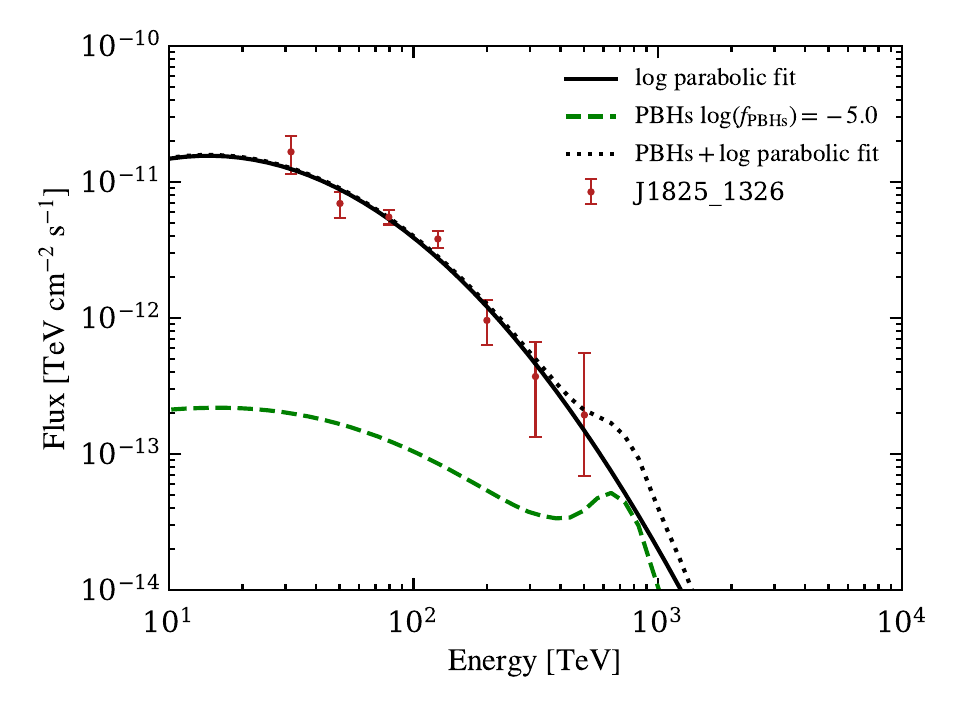}
	\includegraphics[width=0.49\textwidth]{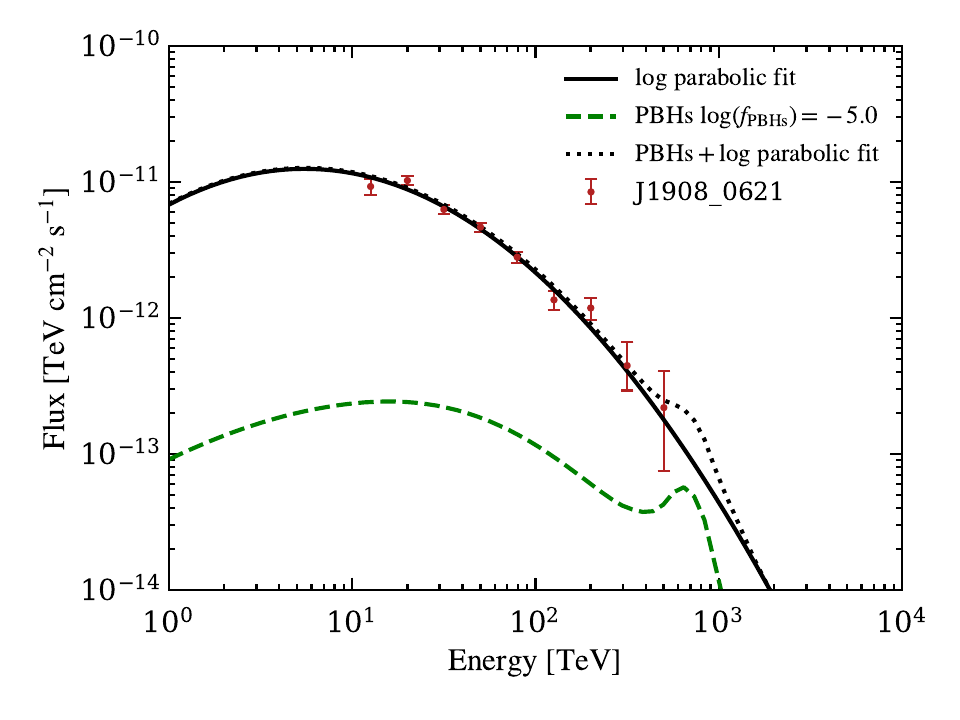}
	\includegraphics[width=0.49\textwidth]{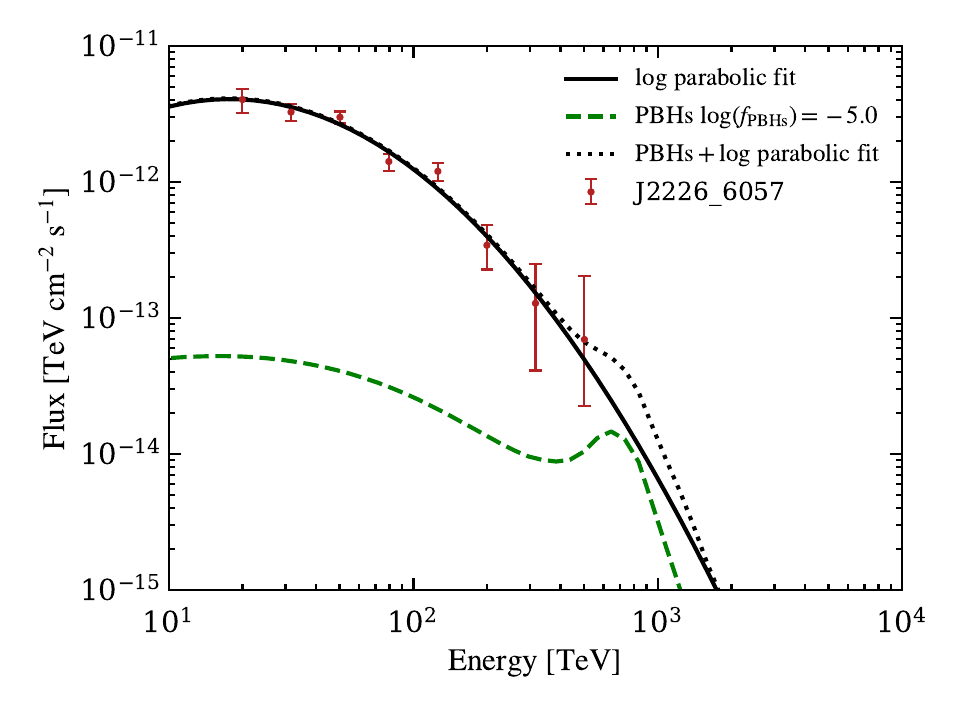}
	\caption{The best-fit spectra for the Crab Nebula, LHAASO J1825-1326, LHAASO J1908+0621, and LHAASO J2226+6057. The solid and dashed lines represent the spectra under the IC (null) and alternative (IC + PBH) hypotheses, respectively. We present the $M_{\rm PBH}=10^8$ g computed for each situation of the Galactic Sources to be an example in green dashed line, with $k=1$. The red points denote the photon spectra measured by LHAASO \cite{LHAASO:2021gok}.}
	\label{fig:spectra_compare_predict}
\end{figure}

\section{Analysis and Results}\label{sec:ana}
As discussed in Section \ref{sec:intro}, we employ observations from the LHAASO of four Galactic sources to derive conservative limits on the parameter space of PBHs affected by the memory burden effect.
First, the observed gamma-ray spectra for these Galactic sources can be modeled by an expected Inverse-Compton (IC) emission spectrum from accelerated electrons within a magnetized nebula. This spectrum is described by a log-parabolic function, given by:
\begin{eqnarray}
	\frac{{\rm d}\phi^{\rm IC}_{\gamma}}{{\rm d}E_\gamma} = F_0 \left(\frac{E}{E_0}\right)^{-a-b\log{(E/E_0)}},
\end{eqnarray}
where $E_0=10$ TeV. 

When consider the contribution from PBHs, the best-fit spectrum can be obtained by minimizing the $\chi^2$ function for consider single or summing up all of the sources, which is 
\begin{eqnarray}
	\chi^2 = \sum_{i,j} \left[\frac{\phi^{\rm predict}_{i,j}(\theta)-\phi^{\rm data}_{i, j}}{\sigma^{\rm data}_{i,j}}\right]^2,
	\label{equ:chi2_1}
\end{eqnarray}
where $\phi^{\rm data}_{i,j}$, $\phi^{\rm predict}_{i, j}$ and $\sigma^{\rm data}_{i,j}$ represent the observed spectrum, predicted spectrum and experimental uncertainty of the photon flux in the $i$-th energy bin, $j$-th source, respectively. The predict spectrum is $\frac{{\rm d}^2\phi_{\gamma}^{\rm GS}}{{\rm d}E_\gamma {\rm d}t} + \frac{{\rm d}\phi^{\rm IC}_{\gamma}}{{\rm d}E_\gamma} $, and its parameter space $\theta=\left\lbrace f_{\rm PBH}, M_{\rm PBH}, k, F_0, a, b \right\rbrace$, which first three are parameters for memory-burdened PBHs and the last three $\left\lbrace F_0, a, b\right\rbrace$ represents the best-fit values for the intrinsic parameters for galactic sources spectra.
Therefore, for each selected PBHs parameters $M_{\rm PBH}$ and $k$, we determine the maximum allowed value of $f_{\rm PBH}$ by evaluating $\Delta \chi^2=-2\ln \mathcal{L}$ and applying Wilks' theorem under four degree of freedom. 

For the Crab Nebula, in addition to the LHAASO observations, we have also compiled data from previous experiments, including HAWC \cite{Abeysekara_2019}, AS$\gamma$ \cite{Liu:2021lxk}, HEGRA \cite{Aharonian_2004}, MAGIC \cite{MAGIC:2007qhf}, TibetIII \cite{Amenomori_2009} and HESS \cite{HESS:2006fka}. To ensure consistency across all experiments, accounting for their respective energy scale uncertainties, we employ a $\chi^2$ minimization technique to determine the best-fit parameters. The $\chi^2$ function is defined as follows:
\begin{eqnarray}
	\chi^2 = \sum_{i,j} \left[\frac{\phi^{\rm predict}_{i,j}(\theta)-f_j^{n-1}\phi^{\rm data}_{i, j}}{f_j^{n-1}\sigma^{\rm data}_{i,j}}\right]^2 + \sum_{j} \left[ \frac{(f_j-1)}{\delta f_j} \right]^2,
	\label{equ:chi2_2}
\end{eqnarray}
where the subscripts $i,j$ correspond to those in eqnarray \ref{equ:chi2_1}. Given that the experimental data are presented in the form of $E^n ({\rm d}N/{\rm d}E)$, the quantities $\phi$ and $\sigma$ are scaled by a factor of $f^{n-1}$, where $n=2$ in this context. This implies that $f_j$ represents an energy scale uncertainty, accounting for significant uncertainties in photon energy reconstruction in these measurements.
Following the methodology outlined in reference \cite{Bi:2020ths}, we adopt the following values for the scaling factor $f$: $1.15$ for HEGRA and HESS, $1.0$ for MAGIC, $0.92$ for Tibet AS$\gamma$, and $0.86$ for HAWC. Additionally, the uncertainties in these scaling factors, denoted by $\delta f$, are $0.15$ for HEGRA and MAGIC, $0.12$ for Tibet AS$\gamma$, and $0.14$ for HAWC.

\begin{figure*}[t]
	\centering
	\includegraphics[width=0.49\textwidth]{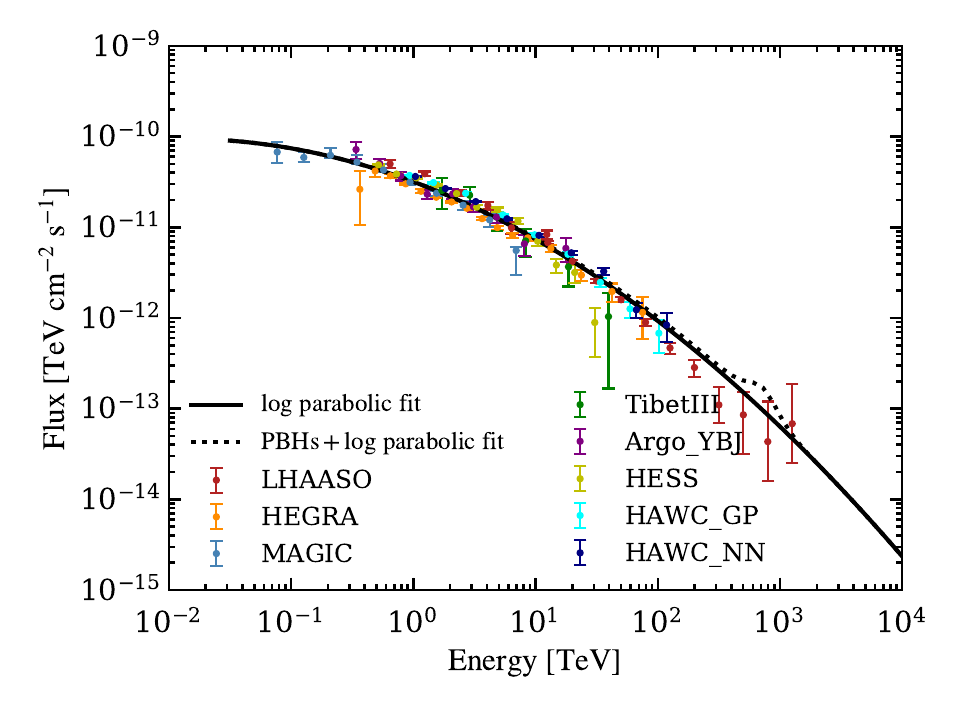}
	\includegraphics[width=0.49\textwidth]{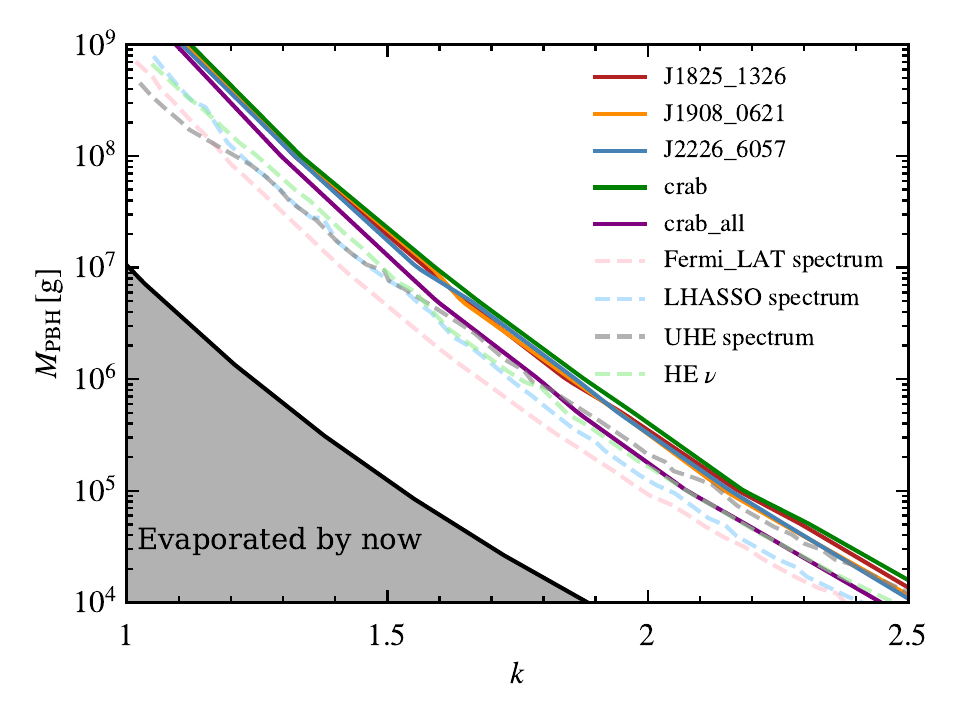}
	\caption{\textit{left} figure refers to the best-fit photon spectra for the Crab Nebula with all the experiments we adapt in this work. The solid and dashed black lines represent the spectra under the IC (null) and alternative (IC + PBH) hypotheses, with $f_{\rm PBH}=10^{-5}$ and $M_{\rm PBH}=10^8$ g, $k=1.0$ for example. \textit{right} figure is the gamma-ray limits placed at 95\% C.L. in the $M_{\rm PBH}-k$ plane with $f_{\rm PBH}=1$, in case of the Galactic Sources LHAASO J1825-1326 (solid red line), LHAASO J1908+0621 (solid orange line), J2226+6057 (solid blue line), Crab Nebula from LHAASO (solid green line) and Crab Nebula from all the experiments we mentioned in the text (solid purple line).}
	\label{fig:kM0&craball}
\end{figure*}

We present the best-fit spectral results of our analysis in figure \ref{fig:spectra_compare_predict}, which includes the IC emission modeled by a log-parabolic function and the scenario where the contribution of PBHs is incorporated. An intuitive modification of the predicted spectrum is observed when PBHs are included.
The left panel of figure \ref{fig:kM0&craball} shows the collected experimental data for the Crab Nebula, compared with the spectra of the log-parabolic fit and the added contribution of PBHs.
The $M_{\rm PBH}-k$ parameter space of memory-burdened PBHs is constrained at the 95\% confidence level (C.L.), as shown on the right panel of figure \ref{fig:kM0&craball}. The gray region indicates the parameter space where PBHs have completely evaporated within cosmological timescales. The colored lines depict the upper limits for different Galactic sources. Here, we fix $f_{\rm PBH}=1$ for the entropy index $k$, implying that the region above the lines corresponds to PBHs that can fully constitute the DM component of the Universe and have not yet evaporated. This plot demonstrates that previous constraints provide slightly tighter limits from Galactic sources than those from gamma-ray diffuse emission or neutrino studies over the range of $k$ from 1 to 2.5.

\begin{figure*}[t]
	\centering
	\includegraphics[width=0.65\textwidth]{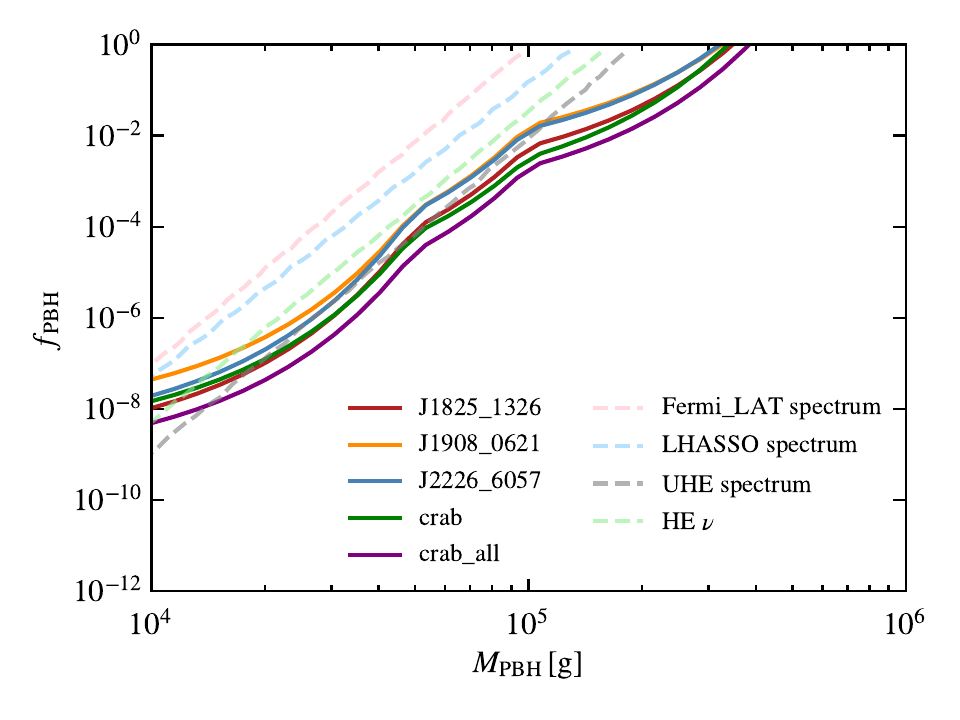}
	\caption{The bounds at the 95\% C.L. in the $f_{\rm PBH}$ and $M_{\rm PBH}$ plane of light PBHs resulting from the LHAASO observations of the Crab Nebula, LHAASO J2226+6057, LHAASO J1908+0621, and LHAASO J1825-1326. The memory-burden parameter $k$ is settle down with $2.0$. Comparing with previous constraint lines from Fermi-LAT for (dashed translucent pink line), LHAASO diffuse spectrum (dashed translucent grey line), UHE diffuse gamma-ray (dashed translucent blue line), and HE neutrinos (dashed translucent green line).}
	\label{fig:const_line}
\end{figure*}
In figure \ref{fig:const_line}, we superimpose the PBHs spectra on the IC emission spectrum using an example with $f_{\rm PBH} =10^{-5}$, $M_{\rm PBH}=10^8{~\rm g}$, and $k=2.0$. By fixing the $k$ value, we derive the bounds on the PBH parameter space for $f_{\rm PBH}$ and $M_{\rm PBH}$, and report the primary constraint lines for PBHs contributing to the Galactic source spectra. For comparison, we also include previous results from high-energy neutrinos (HE$\nu$), the diffuse high-energy gamma-ray spectrum from Fermi-LAT, LHAASO, and Ultra-High Energy (UHE) studies \cite{Chianese:2025wrk}.
For different Galactic sources, the l.o.s distance can affect the contribution of PBHs to high-energy gamma-rays by influencing the range of the integral calculation. Specifically, PBHs located at greater distances can produce more signals, but these signals are also subject to attenuation along their path to the detectors. Consequently, this results in slightly different performances in the constraint lines. Owing to the specific higher energies of these Galactic sources, we obtain more stringent results compared to previous works. As depicted in the figure, the constraints from Galactic sources are tighter than those from previous studies when $M_{\rm PBH} > 10^5 {~\rm g}$, with the collected Crab Nebula data providing the most stringent bounds in these cases, which excludes $M_{\rm PBH}<2\times 10^5{~\rm g}$ at $k=2.0$ from composing all of the DM abundance.

\section{Conclusion}\label{sec:concl}
In this paper, we investigate the high-energy gamma-ray emissions from light PBHs through their evaporation and memory-burden effects. The memory-burden effect has reopened the mass window $<10^{15}$ g for PBHs, rendering light PBHs significant not only in the early Universe but also in the present era. The potential presence of PBHs in the Galactic center could influence the gamma-ray spectra of Galactic sources. Here, we focus on four such sources from the LHAASO experiment: the Crab Nebula, LHAASO J2226+6057, LHAASO J1908+0621, and LHAASO J1825-1326. These sources provide more stringent constraints than those from previous studies of gamma-ray diffuse emission or neutrino emission. While the differences among the four sources are minor, the broader energy spectra from numerous experiments for the Crab Nebula yield the most stringent constraints for $k=2.0$.

We find that ultra-high-energy gamma-ray sources in the Galactic center can impose strong bounds on the parameter space of memory-burdened PBHs, under the IC background log-parabolic fitting. We also anticipate that future multi-messenger data will further improve the constraints on memory-burdened PBHs as a dark matter candidate.

\section*{Acknowledgements}
Y.F.Z. is supported by the National Key R\&D Program of China (Grant No. 2017YFA0402204), the CAS Project for Young Scientists in Basic Research YSBR-006, and the NSFC (Grants No. 11821505, No. 11825506, and No. 12047503).

\normalem
\bibliography{references}

\providecommand{\href}[2]{#2}\begingroup\raggedright\begin{thebibliography}{10}

\bibitem{1975CMaPh..43..199H}
S.~W. {Hawking}, \emph{{Particle creation by black holes}},
  \href{http://dx.doi.org/10.1007/BF02345020}{\emph{Communications in
  Mathematical Physics} {\bf 43} (Aug., 1975) 199--220}.

\bibitem{10.1093/mnras/152.1.75}
S.~Hawking, \emph{{Gravitationally Collapsed Objects of Very Low Mass}},
  \href{http://dx.doi.org/10.1093/mnras/152.1.75}{\emph{Monthly Notices of the
  Royal Astronomical Society} {\bf 152} (04, 1971) 75--78}.

\bibitem{Bambi:2008kx}
C.~Bambi, A.~D. Dolgov and A.~A. Petrov, \emph{{Primordial black holes and the
  observed Galactic 511-keV line}},
  \href{http://dx.doi.org/10.1016/j.physletb.2009.10.053}{\emph{Phys. Lett. B}
  {\bf 670} (2008) 174--178}, [\href{http://arxiv.org/abs/0801.2786}{{\tt
  0801.2786}}]. [Erratum: Phys.Lett.B 681, 504--504 (2009)].

\bibitem{DeRocco:2019fjq}
W.~DeRocco and P.~W. Graham, \emph{{Constraining Primordial Black Hole
  Abundance with the Galactic 511 keV Line}},
  \href{http://dx.doi.org/10.1103/PhysRevLett.123.251102}{\emph{Phys. Rev.
  Lett.} {\bf 123} (2019) 251102}, [\href{http://arxiv.org/abs/1906.07740}{{\tt
  1906.07740}}].

\bibitem{Laha:2019ssq}
R.~Laha, \emph{{Primordial Black Holes as a Dark Matter Candidate Are Severely
  Constrained by the Galactic Center 511 keV $\gamma$ -Ray Line}},
  \href{http://dx.doi.org/10.1103/PhysRevLett.123.251101}{\emph{Phys. Rev.
  Lett.} {\bf 123} (2019) 251101}, [\href{http://arxiv.org/abs/1906.09994}{{\tt
  1906.09994}}].

\bibitem{laha_integral_2020}
R.~Laha, J.~B. Mu{\~n}oz and T.~R. Slatyer, \emph{{{INTEGRAL}} constraints on
  primordial black holes and particle dark matter},
  \href{http://dx.doi.org/10.1103/PhysRevD.101.123514}{\emph{Physical Review D}
  {\bf 101} (June, 2020) 123514}, [\href{http://arxiv.org/abs/2004.00627}{{\tt
  2004.00627}}].

\bibitem{Iguaz:2021irx}
J.~Iguaz, P.~D. Serpico and T.~Siegert, \emph{Isotropic {{X-ray}} bound on
  {{Primordial Black Hole Dark Matter}}},
  \href{http://dx.doi.org/10.1103/PhysRevD.103.103025}{\emph{Physical Review D}
  {\bf 103} (May, 2021) 103025}, [\href{http://arxiv.org/abs/2104.03145}{{\tt
  2104.03145}}].

\bibitem{Chen:2021ngo}
S.~Chen, H.-H. Zhang and G.~Long, \emph{Revisiting the constraints on
  primordial black hole abundance with the isotropic gamma ray background},
  \href{http://dx.doi.org/10.1103/PhysRevD.105.063008}{\emph{Physical Review D}
  {\bf 105} (Mar., 2022) 063008}, [\href{http://arxiv.org/abs/2112.15463}{{\tt
  2112.15463}}].

\bibitem{Tan:2022lbm}
X.-H. Tan, Y.-J. Yan, T.~Qiu and J.-Q. Xia, \emph{{Searching for the Signal of
  a Primordial Black Hole from CMB Lensing and \ensuremath{\gamma}-Ray
  Emissions}},
  \href{http://dx.doi.org/10.3847/2041-8213/ac9668}{\emph{Astrophys. J. Lett.}
  {\bf 939} (2022) L15}, [\href{http://arxiv.org/abs/2209.15222}{{\tt
  2209.15222}}].

\bibitem{Korwar:2023kpy}
M.~Korwar and S.~Profumo, \emph{Updated constraints on primordial black hole
  evaporation},
  \href{http://dx.doi.org/10.1088/1475-7516/2023/05/054}{\emph{Journal of
  Cosmology and Astroparticle Physics} {\bf 2023} (may, 2023) 054}.

\bibitem{Tan:2024nbx}
X.-h. Tan and J.-q. Xia, \emph{{Revisiting bounds on primordial black hole as
  dark matter with X-ray background}},
  \href{http://dx.doi.org/10.1088/1475-7516/2024/09/022}{\emph{JCAP} {\bf 09}
  (2024) 022}, [\href{http://arxiv.org/abs/2404.17119}{{\tt 2404.17119}}].

\bibitem{Huang:2024xap}
J.-Z. Huang and Y.-F. Zhou, \emph{{Constraints on evaporating primordial black
  holes from the AMS-02 positron data}},
  \href{http://dx.doi.org/10.1103/PhysRevD.111.083525}{\emph{Phys. Rev. D} {\bf
  111} (2025) 083525}, [\href{http://arxiv.org/abs/2403.04987}{{\tt
  2403.04987}}].

\bibitem{Dvali:2018xpy}
G.~Dvali, \emph{{A Microscopic Model of Holography: Survival by the Burden of
  Memory}},  \href{http://arxiv.org/abs/1810.02336}{{\tt 1810.02336}}.

\bibitem{Dvali:2020wft}
G.~Dvali, L.~Eisemann, M.~Michel and S.~Zell, \emph{{Black hole metamorphosis
  and stabilization by memory burden}},
  \href{http://dx.doi.org/10.1103/PhysRevD.102.103523}{\emph{Phys. Rev. D} {\bf
  102} (2020) 103523}, [\href{http://arxiv.org/abs/2006.00011}{{\tt
  2006.00011}}].

\bibitem{Dvali:2024hsb}
G.~Dvali, J.~S. Valbuena-Berm\'udez and M.~Zantedeschi, \emph{{Memory burden
  effect in black holes and solitons: Implications for PBH}},
  \href{http://dx.doi.org/10.1103/PhysRevD.110.056029}{\emph{Phys. Rev. D} {\bf
  110} (2024) 056029}, [\href{http://arxiv.org/abs/2405.13117}{{\tt
  2405.13117}}].

\bibitem{Haque:2024eyh}
M.~R. Haque, S.~Maity, D.~Maity and Y.~Mambrini, \emph{{Quantum effects on the
  evaporation of PBHs: contributions to dark matter}},
  \href{http://dx.doi.org/10.1088/1475-7516/2024/07/002}{\emph{JCAP} {\bf 07}
  (2024) 002}, [\href{http://arxiv.org/abs/2404.16815}{{\tt 2404.16815}}].

\bibitem{Dvali:2021tez}
G.~Dvali, O.~Kaikov and J.~S.~V. Berm\'udez, \emph{{How special are black
  holes? Correspondence with objects saturating unitarity bounds in generic
  theories}}, \href{http://dx.doi.org/10.1103/PhysRevD.105.056013}{\emph{Phys.
  Rev. D} {\bf 105} (2022) 056013},
  [\href{http://arxiv.org/abs/2112.00551}{{\tt 2112.00551}}].

\bibitem{Alexandre:2024nuo}
A.~Alexandre, G.~Dvali and E.~Koutsangelas, \emph{{New mass window for
  primordial black holes as dark matter from the memory burden effect}},
  \href{http://dx.doi.org/10.1103/PhysRevD.110.036004}{\emph{Phys. Rev. D} {\bf
  110} (2024) 036004}, [\href{http://arxiv.org/abs/2402.14069}{{\tt
  2402.14069}}].

\bibitem{Thoss:2024hsr}
V.~Thoss, A.~Burkert and K.~Kohri, \emph{{Breakdown of hawking evaporation
  opens new mass window for primordial black holes as dark matter candidate}},
  \href{http://dx.doi.org/10.1093/mnras/stae1098}{\emph{Mon. Not. Roy. Astron.
  Soc.} {\bf 532} (2024) 451--459},
  [\href{http://arxiv.org/abs/2402.17823}{{\tt 2402.17823}}].

\bibitem{Chianese:2024rsn}
M.~Chianese, A.~Boccia, F.~Iocco, G.~Miele and N.~Saviano, \emph{{Light burden
  of memory: Constraining primordial black holes with high-energy neutrinos}},
  \href{http://dx.doi.org/10.1103/PhysRevD.111.063036}{\emph{Phys. Rev. D} {\bf
  111} (2025) 063036}, [\href{http://arxiv.org/abs/2410.07604}{{\tt
  2410.07604}}].

\bibitem{Dvali:2025ktz}
G.~Dvali, M.~Zantedeschi and S.~Zell, \emph{{Transitioning to Memory Burden:
  Detectable Small Primordial Black Holes as Dark Matter}},
  \href{http://arxiv.org/abs/2503.21740}{{\tt 2503.21740}}.

\bibitem{Montefalcone:2025akm}
G.~Montefalcone, D.~Hooper, K.~Freese, C.~Kelso, F.~Kuhnel and P.~Sandick,
  \emph{{Does Memory Burden Open a New Mass Window for Primordial Black Holes
  as Dark Matter?}},  \href{http://arxiv.org/abs/2503.21005}{{\tt 2503.21005}}.

\bibitem{Liu:2025vpz}
T.-C. Liu, B.-Y. Zhu, Y.-F. Liang, X.-S. Hu and E.-W. Liang,
  \emph{{Constraining the parameters of heavy dark matter and memory-burdened
  primordial black holes with DAMPE electron measurements}},
  \href{http://dx.doi.org/10.1016/j.jheap.2025.100375}{\emph{JHEAp} {\bf 47}
  (2025) 100375}, [\href{http://arxiv.org/abs/2503.13192}{{\tt 2503.13192}}].

\bibitem{Zantedeschi:2024ram}
M.~Zantedeschi and L.~Visinelli, \emph{{Ultralight Black Holes as Sources of
  High-Energy Particles}},  \href{http://arxiv.org/abs/2410.07037}{{\tt
  2410.07037}}.

\bibitem{Ma:2022aau}
X.-H. Ma et~al., \emph{{Chapter 1 LHAASO Instruments and Detector technology
  *}}, \href{http://dx.doi.org/10.1088/1674-1137/ac3fa6}{\emph{Chin. Phys. C}
  {\bf 46} (2022) 030001}.

\bibitem{LHAASO:2021gok}
{\scshape LHAASO} collaboration, Z.~Cao et~al., \emph{{Ultrahigh-energy photons
  up to 1.4 petaelectronvolts from 12 $\gamma$-ray Galactic sources}},
  \href{http://dx.doi.org/10.1038/s41586-021-03498-z}{\emph{Nature} {\bf 594}
  (2021) 33--36}.

\bibitem{Li:2024ivs}
J.~Li, X.-J. Bi, L.-Q. Gao, X.~Huang, R.-M. Yao and P.-F. Yin,
  \emph{{Constraints on axion-like particles from the observation of Galactic
  sources by the LHAASO*}},
  \href{http://dx.doi.org/10.1088/1674-1137/ad361e}{\emph{Chin. Phys. C} {\bf
  48} (2024) 065107}, [\href{http://arxiv.org/abs/2401.01829}{{\tt
  2401.01829}}].

\bibitem{Arbey:2019mbc}
A.~Arbey and J.~Auffinger, \emph{{BlackHawk: A public code for calculating the
  Hawking evaporation spectra of any black hole distribution}},
  \href{http://dx.doi.org/10.1140/epjc/s10052-019-7161-1}{\emph{Eur. Phys. J.
  C} {\bf 79} (2019) 693}, [\href{http://arxiv.org/abs/1905.04268}{{\tt
  1905.04268}}].

\bibitem{Arbey:2021mbl}
A.~Arbey and J.~Auffinger, \emph{{Physics Beyond the Standard Model with
  BlackHawk v2.0}},
  \href{http://dx.doi.org/10.1140/epjc/s10052-021-09702-8}{\emph{Eur. Phys. J.
  C} {\bf 81} (2021) 910}, [\href{http://arxiv.org/abs/2108.02737}{{\tt
  2108.02737}}].

\bibitem{Bauer:2020jay}
C.~W. Bauer, N.~L. Rodd and B.~R. Webber, \emph{{Dark matter spectra from the
  electroweak to the Planck scale}},
  \href{http://dx.doi.org/10.1007/JHEP06(2021)121}{\emph{JHEP} {\bf 06} (2021)
  121}, [\href{http://arxiv.org/abs/2007.15001}{{\tt 2007.15001}}].

\bibitem{Navarro_1997}
J.~F. Navarro, C.~S. Frenk and S.~D.~M. White, \emph{A universal density
  profile from hierarchical clustering},
  \href{http://dx.doi.org/10.1086/304888}{\emph{The Astrophysical Journal} {\bf
  490} (dec, 1997) 493}.

\bibitem{Abeysekara_2019}
A.~U. Abeysekara, A.~Albert, R.~Alfaro, C.~Alvarez, J.~D. Álvarez, J.~R.~A.
  Camacho et~al., \emph{Measurement of the crab nebula spectrum past 100 tev
  with hawc}, \href{http://dx.doi.org/10.3847/1538-4357/ab2f7d}{\emph{The
  Astrophysical Journal} {\bf 881} (aug, 2019) 134}.

\bibitem{Liu:2021lxk}
R.-Y. Liu and X.-Y. Wang, \emph{{Origin of Galactic Sub-PeV Diffuse Gamma-Ray
  Emission: Constraints from High-energy Neutrino Observations}},
  \href{http://dx.doi.org/10.3847/2041-8213/ac02c5}{\emph{Astrophys. J. Lett.}
  {\bf 914} (2021) L7}, [\href{http://arxiv.org/abs/2104.05609}{{\tt
  2104.05609}}].

\bibitem{Aharonian_2004}
F.~Aharonian, A.~Akhperjanian, M.~Beilicke, K.~Bernlöhr, H.-G. Börst,
  H.~Bojahr et~al., \emph{The crab nebula and pulsar between 500 gev and 80
  tev: Observations with the hegra stereoscopic air cerenkov telescopes},
  \href{http://dx.doi.org/10.1086/423931}{\emph{The Astrophysical Journal} {\bf
  614} (oct, 2004) 897}.

\bibitem{MAGIC:2007qhf}
{\scshape MAGIC} collaboration, J.~Albert et~al., \emph{{VHE Gamma-Ray
  Observation of the Crab Nebula and Pulsar with MAGIC}},
  \href{http://dx.doi.org/10.1086/525270}{\emph{Astrophys. J.} {\bf 674} (2008)
  1037--1055}, [\href{http://arxiv.org/abs/0705.3244}{{\tt 0705.3244}}].

\bibitem{Amenomori_2009}
M.~Amenomori, X.~J. Bi, D.~Chen, S.~W. Cui, Danzengluobu, L.~K. Ding et~al.,
  \emph{Multi-tev gamma-ray observation from the crab nebula using the
  tibet-iii air shower array finely tuned by the cosmic ray moon's shadow},
  \href{http://dx.doi.org/10.1088/0004-637X/692/1/61}{\emph{The Astrophysical
  Journal} {\bf 692} (feb, 2009) 61}.

\bibitem{HESS:2006fka}
{\scshape H.E.S.S.} collaboration, F.~Aharonian et~al., \emph{{Observations of
  the Crab Nebula with H.E.S.S}},
  \href{http://dx.doi.org/10.1051/0004-6361:20065351}{\emph{Astron. Astrophys.}
  {\bf 457} (2006) 899--915},
  [\href{http://arxiv.org/abs/astro-ph/0607333}{{\tt astro-ph/0607333}}].

\bibitem{Bi:2020ths}
X.-J. Bi, Y.~Gao, J.~Guo, N.~Houston, T.~Li, F.~Xu et~al., \emph{{Axion and
  dark photon limits from Crab Nebula high energy gamma-rays}},
  \href{http://dx.doi.org/10.1103/PhysRevD.103.043018}{\emph{Phys. Rev. D} {\bf
  103} (2021) 043018}, [\href{http://arxiv.org/abs/2002.01796}{{\tt
  2002.01796}}].

\bibitem{Chianese:2025wrk}
M.~Chianese, \emph{{High-energy gamma-ray emission from memory-burdened
  primordial black holes}},  \href{http://arxiv.org/abs/2504.03838}{{\tt
  2504.03838}}.

\end{thebibliography}\endgroup

\end{document}